\documentclass[]{pasj01}

\Received{$\langle$reception date$\rangle$}
\Accepted{$\langle$acception date$\rangle$}
\Published{$\langle$publication date$\rangle$}
\begin{document}

\title{Interaction between Molecular Clouds and MeV--TeV Cosmic-ray
	Protons Escaped from Supernova Remnants} \author{Ken
	\textsc{Makino},\altaffilmark{1}$^{*}$ Yutaka
	\textsc{Fujita},\altaffilmark{1}$^{*}$ Kumiko K.
	\textsc{Nobukawa},\altaffilmark{2} Hironori \textsc{Matsumoto},\altaffilmark{1,3} Yutaka \textsc{Ohira}\altaffilmark{4} }
	\altaffiltext{1}{Department of Earth and Space Science, Graduate
	School of Science, Osaka University, Toyonaka, Osaka 560-0043,
	Japan} \altaffiltext{2}{Department of Physics, Faculty of
	Science, Nara Women’s University, Kitauoyanishi-machi, 
	Nara 630-8506, Japan}
	\altaffiltext{3}{Project Research Center for Fundamental Sciences, Graduate School of Science, Osaka University
1-1 Machikanetama-cho, Toyonaka, Osaka 560-0043, Japan}
	\altaffiltext{4}{Department of Earth and
	Planetary Science, The University of Tokyo, 7-3-1 Hongo,
	Bunkyo-ku, Tokyo 113-0033, Japan} \email{makino@astro-osaka.jp,
	fujita@astro-osaka.jp}


\maketitle

\begin{abstract}
Recent discovery of the X-ray neutral iron line (Fe~{\small I} K$\alpha$
at 6.40~keV) around several supernova remnants (SNRs) show that MeV
cosmic-ray (CR) protons are distributed around the SNRs and are
interacting with neutral gas there. We propose that these MeV CRs are
the ones that have been accelerated at the SNRs together with GeV--TeV
CRs. In our analytical model, the MeV CRs are still confined in the SNR
when the SNR collides with molecular clouds. After the collision, the
MeV CRs leak into the clouds and produce the neutral iron line
emissions. On the other hand, GeV--TeV CRs had already escaped from the
SNRs and emit gamma-rays through interaction with molecular clouds
surrounding the SNRs. We apply this model to the SNRs W28 and W44 and
show that it can reproduce the observations of the iron line intensities
and the gamma-ray spectra. This can be another support of a hadronic
scenario for the gamma-ray emissions from these SNRs.
\end{abstract}

\section{Introduction}

Supernova remnants (SNRs) have been thought to be the site where
cosmic-rays (CRs) with an energy of $E\lesssim 10^{15.5}$~eV (the knee
energy) are accelerated. The most plausible process of the CR
acceleration is a diffusive shock acceleration (DSA) at their shock
front
\citep{1978MNRAS.182..147B,1978ApJ...221L..29B,1983RPPh...46..973D}. An
excess in GeV-TeV gamma-rays have been observed for SNRs associated with
molecular clouds (e.g. \cite{2009ApJ...706L...1A,2008A&A...481..401A}),
which is believed to be evidence that CRs are actually accelerated at
SNRs. However, there has been a debate on whether the origin of the
gamma-rays is leptonic or hadronic.  If it is leptonic, these signals
could be caused by bremsstrahlung or inverse Compton scattering of
electrons. However, recent detections of the characteristic pion-decay
feature in the gamma-ray spectra strongly suggest that the origin should
be hadronic
(e.g. \cite{2013Sci...339..807A,2016ApJ...816..100J}). However, since
gamma-rays are produced only by CR protons with energies of $E\gtsim
$~GeV, the gamma-ray observations cannot probe lower-energy ($\sim
$~MeV) protons that are likely to be accelerated at the same time. If
the existence of the lower-energy protons are confirmed, it could be a
further support of the hadronic scenario.

The lower-energy protons could be studied through ionization signatures
in molecular gas, because those protons are very efficient in ionizing
molecular gas \citep{2012A&A...541A.126S,2015ICRC...34..518K}. Alternatively, they could be
investigated through X-ray neutral iron line emissions (Fe~{\small I}
K$\alpha$ at 6.40~keV). Recently, \citet{2018ApJ...854...87N} actually
detected the iron line emissions from five SNRs interacting with
molecular clouds from {\em Suzaku} archive data. However, the line
emissions could be produced not only by low-energy protons but also
electrons and X-rays.  \citet{2018ApJ...854...87N} concluded that
protons are most likely because of the observed large equivalent width
of the line and non-existence of nearby X-ray sources. The iron line
emissions were also reported by \citet{2018PASJ...70...35O} for the SNR
W28 (see also
\cite{2014PASJ...66..124S,2016PASJ...68S...8S,2018PASJ...70...23S}).

In the hadronic model, two scenarios have been considered for the
gamma-ray emissions associated with molecular clouds. One is the direct
interaction scenario, in which an SNR directly interacts with molecular
clouds (e.g. \cite{2000ApJ...538..203B,2015ApJ...806...71L}).  In
particular, reacceleration and/or compression of Galactic background CR
protons may boost their energy and create gamma-ray emissions from
molecular clouds
\citep{2010ApJ...723L.122U,2014ApJ...784L..35T,2016A&A...595A..58C,2019MNRAS.482.3843T}. However,
this scenario could face difficulty in explaining the neutral iron line
emissions from MeV protons. This is because the ionization cooling time
of the low-energy protons is very short in high-density molecular clouds
and thus it is unlikely that the clouds contain those protons as
background particles. Thus, in this study we focus on another scenario
called the escaping scenario, in which the molecular clouds passively
interact with the CR particles escaping from an adjacent SNR
(e.g. \cite{1996A&A...309..917A,2009ApJ...707L.179F,2009MNRAS.396.1629G,2010MNRAS.409L..35L}). 
We aim to explain both the gamma-ray and neutral iron line emissions
based on the escaping scenario for the first time.

This paper is organized as follows. In section~\ref{sec:model}, we
explain our model about the escape of CR protons from SNRs and
interaction between the CRs and molecular clouds. In
section~\ref{sec:results}, we apply our model to two SNRs (W28 and W44)
and show that both neutral iron line emissions and gamma-ray spectra can
be explained. The results are discussed in
section~\ref{sec:disc}. Conclusions are given in
section~\ref{sec:conc}. Hereafter, we refer to CR protons as CRs.

\section{Models}
\label{sec:model}

\subsection{Distribution of high-energy CRs escaped from an SNR}
\label{sec:escape}

In this subsection, we summarize the derivation of the distribution
function of high-energy ($\gtrsim$~GeV) CRs escaped from an SNR based on
the model given by \citet{2011MNRAS.410.1577O}. 

We solve a diffusion equation
\begin{equation}
\label{eq:diff}
\frac{\partial f}{\partial t}(t,\mbox{\boldmath $r$},p) 
- D_{\rm ISM}(p) \Delta f(t,\mbox{\boldmath $r$},p) 
= q_{s}(t,\mbox{\boldmath $r$},p)\:,
\end{equation}
where $\mbox{\boldmath $r$}$ is the position, $p$ is the CR momentum,
$f(t,\mbox{\boldmath $r$},p)$ is the distribution function, $D_{\rm
ISM}(p)$ is the diffusion coefficient in the interstellar medium (ISM)
around the SNR, and $q_{\rm s} (t,\mbox{\boldmath $r$},p)$ is the source
term of CRs.

In the following, we assume that the SNR is spherically symmetric and
$r$ is the distance from the SNR center. Moreover, we assume that CRs
with a momentum $p$ escape from the SNR at $t=t_{\rm esc}(p)$
\citep{2005A&A...429..755P,2010A&A...513A..17O}. In the case of a point
source, the source term is give by $q_{\rm s} = N_{\rm
esc}(p)\delta(\mbox{\boldmath $r$} )\delta[t-t_{\rm esc}(p)]$, and the
solution is
\begin{equation}
\label{eq:fpoint}
 f_{\rm point}(t,r,p) 
= \frac{\exp[-(r/R_{\rm d})^2]}{\pi^{3/2}R_{\rm d}^3}N_{\rm esc}(p)\:,
\end{equation}
where
\begin{equation}
 R_{\rm d}(t,p) = \sqrt{4\: D_{\rm ISM}(p)[t-t_{\rm esc}(p)]}\:,
\end{equation}
and 
\begin{equation}
 N_{\rm esc}(p)=\int dt\int d^3\mbox{\boldmath $r$}\: q_{\rm s} (t,r,p)
\:,
\end{equation}
which is the spectrum of the whole escaped CRs.

In realty, CRs escape from the surface of the SNR, $R_{\rm esc}(p)$, and
the source term should be,
\begin{equation}
\label{eq:qsSNR}
q_s(p) = \frac{N_{{\rm esc}}(p)}{4\pi r^2} 
\delta[r-R_{{\rm esc}}(p)]\delta[t-t_{{\rm esc}}(p)]\:.
\end{equation}
For this source term, the solution of equation~(\ref{eq:diff}) can be
derived using equation~(\ref{eq:fpoint}) as the Green function:
\begin{eqnarray}
\label{eq:fext}
f(t,r,p) &=& \int d^3\mbox{\boldmath $r$}' f_{\rm point}(t,|r-r'|,p) 
\frac{\delta[r'-R_{\rm esc}(p)]}{4\pi r'^2}\nonumber\\ 
 &=& \frac{ e^{-(\frac{r-R_{{\rm esc}}(p)}{R_{\rm d}(t,p)})^2 } 
- e^{-(\frac{r+R_{{\rm esc}}(p)}{R_{\rm d}(t,p)})^2 } }
{4\pi^{3/2}R_{\rm d}(t,p)R_{{\rm esc}}(p)r  } N_{{\rm esc}}(p)\:.
\end{eqnarray}

We need to specify the escape time $t_{\rm esc}(p)$, the radius $R_{\rm
esc}(p)$, and the spectrum $N_{\rm esc}(p)$. We assume that the SNR is
in the Sedov phase and CRs are accelerated through a DSA. Thus, CRs are
scattered back and forth across the shock front by magnetic turbulence
during the acceleration. The diffusion coefficient around the shock,
$D_{\rm sh}(p)$, is expected to be much smaller than $D_{\rm ISM}(p)$
for a given $p$, and the diffusion length of the CRs is $\sim D_{\rm
sh}(p)/u_{\rm sh}$, where $u_{\rm sh}$ is the velocity of the shock
front. We assume that if the CRs cross an escape boundary outside the
shock front, they escape from the SNR. Thus, the momentum of escaping
CR, $p_{\rm esc}$, is given by
\begin{equation}
\label{eq:escape}
\frac{D_{\rm sh}(p_{\rm esc})}{u_{\rm sh}} \sim l_{\rm esc}\:,
\end{equation}
where $l_{\rm esc}$ is the distance of the escape boundary from the shock
front. We adopt a geometrical confinement condition $l_{\rm esc}=\kappa
R_{\rm sh}$ and assume that $\kappa=0.04$
\citep{2005A&A...429..755P,2010A&A...513A..17O}. The escape momentum
$p_{\rm esc}$ is expected to be a decreasing function of the shock
radius. Here, we adopt a phenomenological power-law relation:
\begin{equation}
\label{eq:pesc}
p_{\rm esc} = p_{\rm max}
\left(\frac{ R_{\rm sh} }{ R_{\rm Sedov} } \right)^{-\alpha}\:,
\end{equation}
where $p_{\rm max}$ and $R_{\rm Sedov}$ are the escape momentum and the
shock radius at the beginning of the Sedov phase ($t=t_{\rm Sedov}$),
respectively. Following \citet{2011MNRAS.410.1577O}, we assume that the
index is $\alpha=6.5$, which well reproduces gamma-ray spectra of
SNRs. While \citet{2011MNRAS.410.1577O} assumed that $p_{\rm max}c =
10^{15.5}$~eV (the knee energy), we assume that $p_{\rm max}c <
10^{15.5}$~eV and treat it as a parameter because there has been no
direct evidence that CRs are accelerated up to the knee energy at SNRs
(e.g. \cite{2017AIPC.1792b0002G}).

Since we assumed that the SNR is in the Sedov phase, the shock
radius is represented by
\begin{equation}
\label{eq:Rsh}
 R_{\rm sh}(t) 
= R_{\rm Sedov}\left(\frac{t}{t_{\rm Sedov}}\right)^{2/5}\:,
\end{equation}
and the escaping radius is given by
\begin{equation}
\label{eq:Resc}
 R_{\rm esc}(t) = (1+\kappa)R_{\rm sh}(t)\:.
\end{equation}
We assume that $R_{\rm Sedov}=2.1$~pc and $t_{\rm Sedov}=210$~yr
following \citet{2011MNRAS.410.1577O}. Eliminating $R_{\rm sh}$ from
equations~(\ref{eq:pesc}) and (\ref{eq:Rsh}) and replacing $p_{\rm esc}$
and $t$ with $p$ and $t_{\rm esc}$, respectively, we obtain
\begin{equation}
\label{eq:tesc}
 t_{\rm esc}(p) = t_{\rm Sedov}\left(\frac{p}{p_{\rm max}}\right)
^{-5/(2\alpha)}\:.
\end{equation}

We assume that the CR spectrum at the shock front is always represented
by a single power-law $\propto p^{-s}$ and the number of CRs in the
momentum range $(m_{\rm p} c, m_{\rm p} c + dp)$ in the SNR is $K(R_{\rm
sh}) dp \propto R_{\rm sh}^{\beta}$, where $m_{\rm p}$ is the proton
mass. The factor $K(R_{\rm sh})$ corresponds to the normalization of the
CR spectrum confined in the SNR.  If we assume a thermal leakage model
for CR injection, the index is $\beta=3(3-s)/2$
\citep{2010A&A...513A..17O}. Based on these assumptions, the spectrum of
the escaped CRs ($p>p_{\rm esc}$) is written as
\begin{equation}
\label{eq:Nesc}
 N_{\rm esc}(p)\propto p^{-(s+\beta/\alpha)}\:,
\end{equation}
\citep{2010A&A...513A..17O}. Note that the spectrum of the whole escaped
CRs ($p>p_{\rm esc}$) is represented by $\propto p^{-(s+\beta/\alpha)}$
regardless of time [see equation~(\ref{eq:qsSNR})]. We determine the
normalization of equation~(\ref{eq:Nesc}) from the total energy of the
escaped CRs with $pc>1$~GeV ($E_{\rm tot,CR}$), which is treated as a
parameter.

For the diffusion coefficient in the ISM, we assume the following form,
\begin{equation}
\label{eq:DISM}
D_{\rm ISM}(p) = 10^{28}\:\chi \left(\frac{pc}{10\:\rm GeV}\right)^\delta
\rm cm^{2} s^{-1}\:
\end{equation}
\citep{2011MNRAS.410.1577O}. In this study, we assume
Kolmogorov-type turbulence ($\delta = 1/3$), which is theoretically
motivated and close to the values estimated based on recent observations
($\delta \sim 0.4$; \cite{2015JCAP...12..039E,2015A&A...580A...9G}).
The constant $\chi (\leq 1)$ is introduced because the coefficient
around SNRs can be reduced by waves generated through the stream of
escaping CRs (e.g. \cite{2010ApJ...712L.153F,2011MNRAS.415.3434F}). In
this study, we fix it at $\chi=0.5$.\footnote{The diffusion
coefficient is related to magnetic fluctuations $\delta B$ as in $D_{\rm
ISM}/D_{\rm Bohm}\sim (B/\delta B)^2$, where $D_{\rm Bohm}=(1/3) r_{\rm
L} v_{\rm CR}$ is the Bohm diffusion coefficient, $r_{\rm L}$ is the
gyro-radius, $v_{\rm CR}$ is the velocity of the particle, and $B$ is
the background magnetic field
(e.g. \cite{2016APh....73....1R}). Equation~(\ref{eq:DISM}) indicates
that $\delta B/B\sim 0.002$ for $B=3\rm\: \mu G$ and $pc\sim 1$~GeV.}

\subsection{Low-energy CRs interacting with molecular clouds and iron line emissions}
\label{sec:MeV}

The 6.4~keV neutral iron line emissions have been observed only in the
vicinity of SNRs \citep{2018ApJ...854...87N}. Thus, MeV CRs responsible
for the line emissions are distributed there. Some of the SNRs show a
sign of interaction with molecular clouds through maser emissions
(e.g. \cite{1974A&A....35..153P,1981ApJ...245..105W,1997ApJ...489..143C}). Equation~(\ref{eq:tesc})
shows that MeV CRs escape from an SNR after GeV CRs escape. For the SNRs
we study in section~\ref{sec:results} (W28 and W44),
\citet{2011MNRAS.410.1577O} indicated that while CRs with $E\gtrsim$~GeV
have already escaped from the SNRs at this time, MeV CRs have not. In
the following, we assume that MeV CRs are still confined around the SNRs
when the SNRs contact with the molecular clouds from which the iron line
emissions are detected.

\begin{figure}
 \begin{center}
  \includegraphics[width=8.0cm]{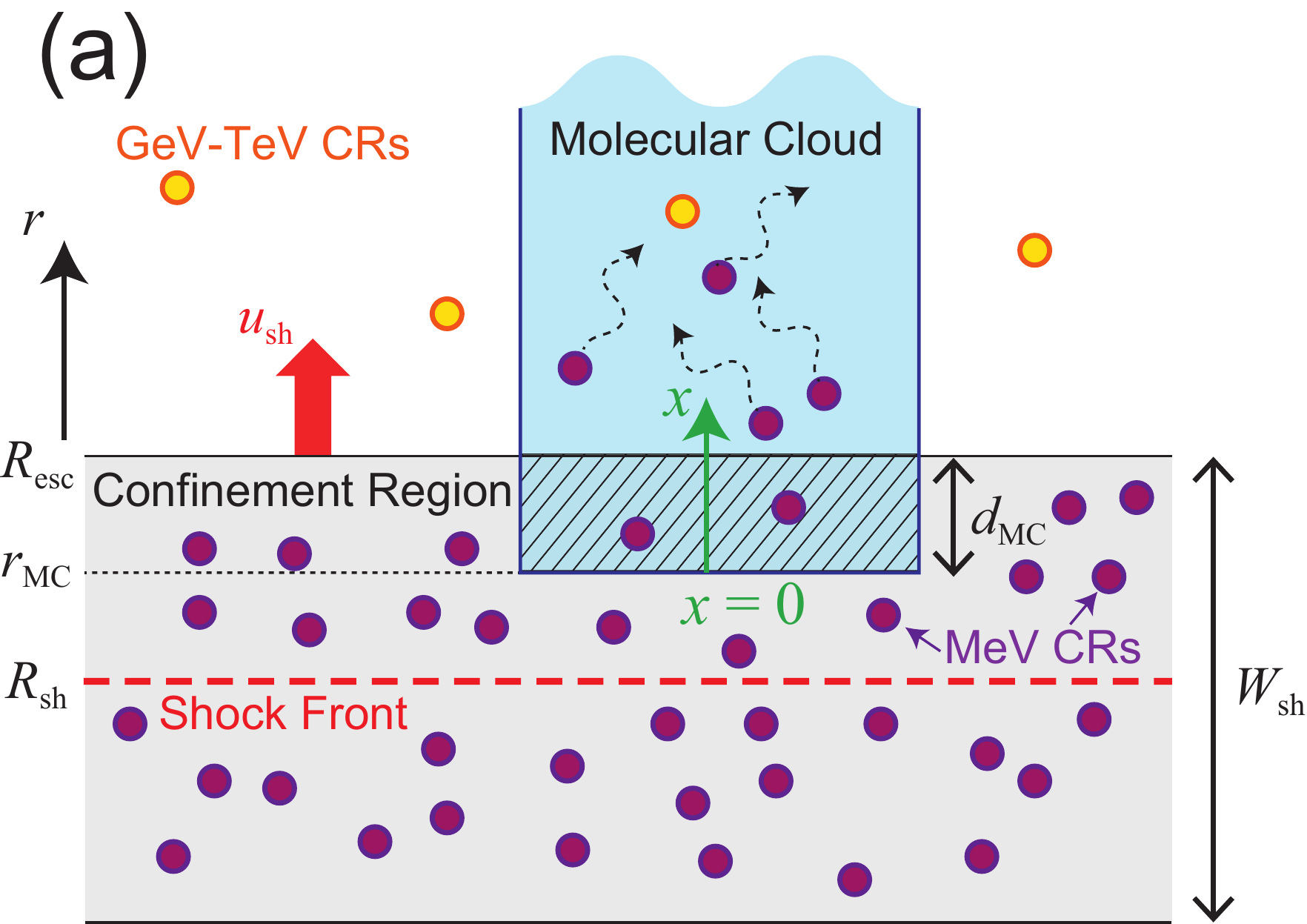}
  \includegraphics[width=8.0cm]{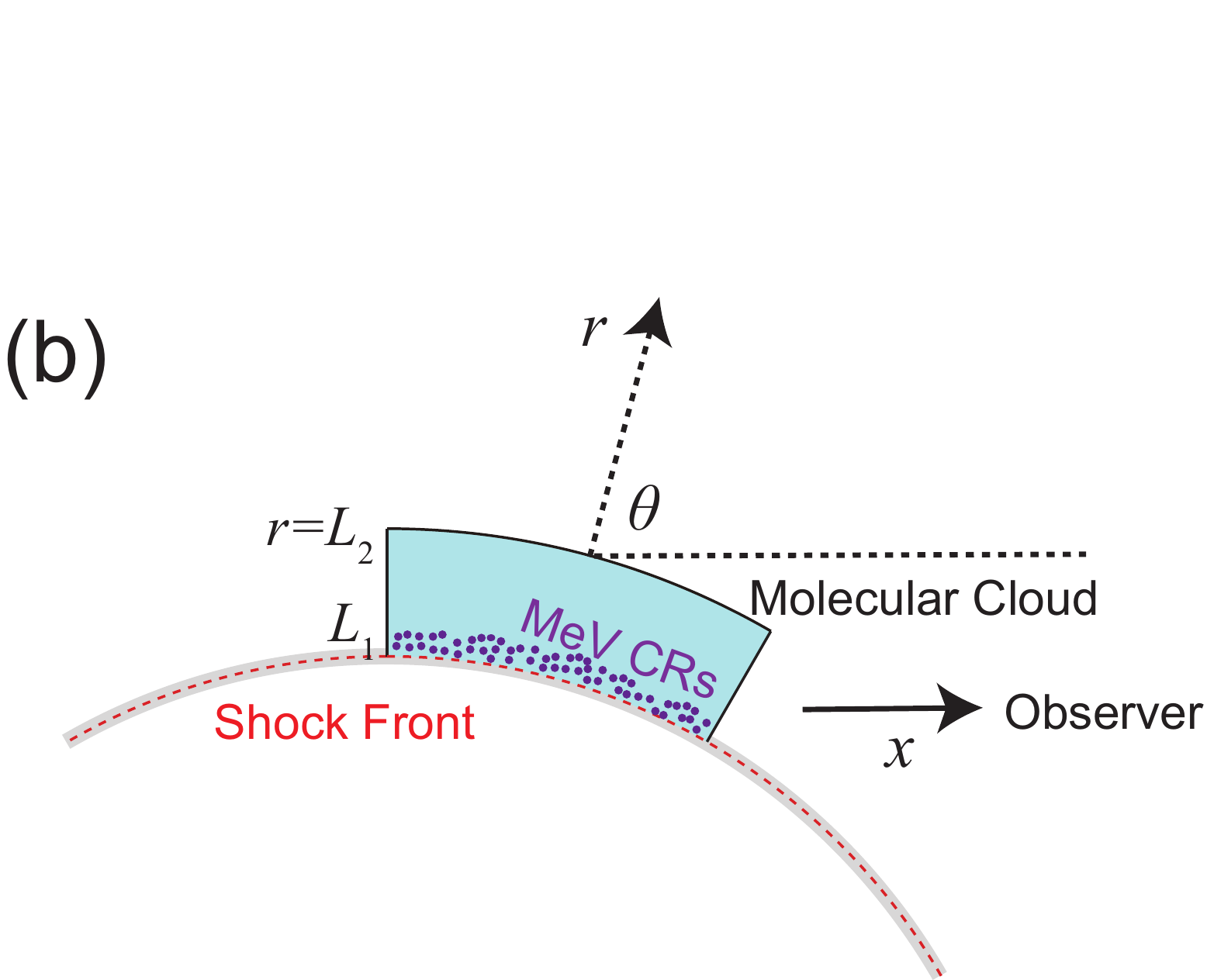}
 \end{center}
 \caption{(a) Schematic figure showing the interaction between the SNR
 and the molecular cloud. MeV CRs seep into the cloud. The directions of
 $r$ and $x$ are parallel ($\theta=0$). The depth of the overlapped
 region is given by $d_{\rm MC}$, Note that the relative position of the
 confinement region to the molecular cloud changes as the shock front
 and the confinement region movies outward with the velocity of $u_{\rm
 sh}$. (b) Zoom-out view of (a), but the directions of $r$ and $x$ are
 not parallel ($\theta\neq 0$).}\label{fig:MeVCRs}
\end{figure}

The spectrum of the low-energy CRs confined in an SNR is written as
\begin{equation}
\label{eq:Nsh}
N_{\rm sh}(t,p) = N_{\rm esc}(p_{\rm esc}(t)) 
\left( \frac{ p }{ p_{\rm esc}(t) } \right)^{-s}\:,
\end{equation}
which is defined for $p< p_{\rm esc}(t)$.  We assume that the CRs are
confined in a region around the shock front with a width of $W_{\rm
sh}\equiv 2\: l_{\rm esc}$ (figure~\ref{fig:MeVCRs}a).  The number
density of the confined CRs is
\begin{equation}
\label{eq:nCRsh}
n_{\rm CR,sh}(t,p) = \frac{N_{\rm sh}(t,p)}{V_{\rm c}}\:,
\end{equation}
where $V_{\rm c}\approx 4\pi R_{\rm sh}^2 W_{\rm sh}$ is the volume of
the confinement region. Note that we do not consider CRs advected into a
far downstream region of the shock front ($r<R_{\rm sh}$) because they
probably lose their energy through adiabatic cooling.

For the sake of simplicity, we here assume that the shock front is a
plane and the molecular cloud is an uniform cuboid, and that the
distribution of CRs in the cloud is one-dimensional
(figure~\ref{fig:MeVCRs}a). We assume that the confined CRs start
seeping into the cloud when the escaping boundary ($r=R_{\rm esc}$)
contacts the surface of a molecular cloud ($r=r_{\rm MC}$). This is
because the CR diffusion coefficient in the cloud is expected be much
larger than that in the confinement region due to the wave dumping
through collisions between protons and neutral particles
\citep{1971ApL.....8..189K}. The CRs are continuously leaked into the
cloud at a rate of $n_{\rm CR,sh}u_{\rm sh}$ per unit area of the shock
front until the confinement region passes the surface of the cloud
($R_{\rm esc}-W_{\rm sh}=r_{\rm MC}$).

The photon number intensity of the neutral iron line is given by
\begin{equation}
\label{eq:line formula}
I_{{\rm 6.4keV}}= \frac{1}{4\pi}\int dE\: \sigma_{\rm 6.4keV}(E) 
v_{\rm CR}(E) n_{\rm H} \int dx \: n_{{\rm CR}}(E,x) \:,
\end{equation}
where $E$, $\sigma_{\rm 6,4keV}(E), v_{\rm CR}(E), n_{\rm H}, n_{\rm
CR}(E,x)$ are the kinetic energy of the CRs, the cross section to
produce the iron line at 6.4~keV, the CR velocity, the number density of
hydrogens in the molecular cloud, and the CR density in the cloud,
respectively. The depth of the cloud in the direction of line of sight
is represented by $x$. In figure~\ref{fig:MeVCRs}a, we assume that the
angle between $x$-direction and $r$-direction is zero ($\theta=0$), and
$x=0$ corresponds to $r=r_{\rm MC}$. For the the cross-section
$\sigma_{{\rm 6,4keV}}(E)$, we use the one for the solar metallicity and
$1<E<10^4$~MeV calculated by \citet{2012A&A...546A..88T}.  We assume
that $\sigma_{\rm 6.4keV}(E) = 0$ for $E>10^4$~MeV and $E<1$~MeV, which
does not affect the results.

If the injection and the cooling of CRs are balanced in the cloud, the
column density of the CRs is written as $n_{\rm CR,sh} u_{\rm sh} t_{\rm
cool}$, where $t_{\rm cool}$ is the cooling time of the CRs and is
calculated using the ionization loss rate given by
\citet{1994A&A...286..983M}. The column density corresponds to the
second integral of equation~(\ref{eq:line formula}). Thus, the line
intensity is represented by
\begin{eqnarray}
\label{eq:line}
I_{\rm 6.4keV}& = &\frac{1}{4\pi}\int dE\: [\sigma_{\rm 6.4keV}(E) 
v_{\rm CR}(E) n_{\rm H}\nonumber \\
&  &  \times
n_{\rm CR,sh}(t,E) u_{\rm sh} t_{\rm cool}(E)]  \:.
\end{eqnarray}
This equation is correct if the injection of CRs into the molecular
cloud is endless. However, the width of the confinement region $W_{\rm
sh}$ is finite and the CR column density $n_{\rm CR,sh} u_{\rm sh}
t_{\rm cool}$ cannot be larger than $n_{\rm CR,sh} W_{\rm sh}$. Since we
do not know how deep the confinement region is immersed in the cloud at
present, we simply assume that the region is half immersed ($d_{\rm
MC}=W_{\rm sh}/2$ in figure~\ref{fig:MeVCRs}a). Thus,
equation~(\ref{eq:line}) is modified as
\begin{eqnarray}
\label{eq:line2}
I_{\rm 6.4keV}& = &\frac{1}{4\pi}\int dE\: [\sigma_{\rm 6.4keV}(E) 
v_{\rm CR}(E) n_{\rm H}\nonumber \\
&  &  \times
n_{\rm CR,sh}(t,E) u_{\rm sh} t'_{\rm cool}(E)]  \:.
\end{eqnarray}
where $t'_{\rm cool}(E)=\min[t_{\rm cool}(E),0.5\: t_{\rm pass}]$ and
$t_{\rm pass}\equiv W_{\rm sh}/u_{\rm sh}$ is the time scale in which
the confinement region passes the surface of the molecular cloud. This
means that the CRs that were originally in the overlapped region (the
shaded region in figure~\ref{fig:MeVCRs}a) when $d_{\rm MC}=W_{\rm
sh}/2$ was satisfied have escaped into the cloud.

However, this correction is not significant. The time scale in which the
confinement region passes the surface of the molecular cloud ($t_{\rm
pass}$) is $\sim 20$\% of the age of the SNR ($\sim 10^4$~yr) for the
parameters we choose in section~\ref{sec:results}. While this time scale
is too short to change the gamma-ray spectrum produced by $\gtrsim$~GeV
CRs, it is much larger than the cooling time of MeV CRs. For example,
CRs with $E\sim 10$~MeV are most effective to create the iron line
emissions due to a large $\sigma_{\rm 6.4keV}(E)$; they have a cooing
time of $t_{\rm cool}\lesssim 100$~yr for $n_{\rm H}>1000\rm\: cm^{-3}$,
which means that $t'_{\rm cool}= t_{\rm cool}$ at $E\sim 10$~MeV.

For the sake of simplicity, we assume that CRs that have entered clouds
do not escape from the clouds before they lose their energy through the
rapid cooling. This may be realized if magnetic fields are oriented in
the clouds so that they trap the CRs. In other words, the iron line
emissions are produced only in the clouds where the fields are properly
distributed. Under this assumption, the intensity represented by
equation~(\ref{eq:line2}) does not depend on the details of the
molecular could such as the total length in the direction of $x$ or the
CR diffusion coefficient inside it.  If the viewing angle $\theta$ is
not zero (figure~\ref{fig:MeVCRs}b), we expect that $I_{\rm 6.4keV}$
becomes larger and $I_{\rm 6.4keV}(\theta)\sim I_{\rm
6.4keV}(\theta=0)/\cos\theta$. From now on, we assume that $\theta$ is
not too close to $90^\circ$ unless otherwise mentioned.

\subsection{Gamma-ray emissions from molecular clouds}

Gamma-rays are produced through $pp$-interaction between CRs and
hydrogens in molecular clouds. We assume that molecular clouds with
density $n_{\rm H}$ are distributed in a shell region between
$r=L_1\approx R_{\rm esc}$ and $r=L_2$ with a filling factor $f_{\rm
gas}$ (figure~\ref{fig:MeVCRs}b). 
The current time $t_{\rm obs}$ is given by 
\begin{equation}
\label{eq:tobs}
 R_{\rm esc}(t_{\rm obs})=L_1\:,
\end{equation}
and equations~(\ref{eq:Rsh}) and (\ref{eq:Resc})\footnote{We implicitly
assumed that the current time is given by $R_{\rm sh}(t_{\rm obs})= L_1$
when we calculate the neutral iron line intensity
[equation~(\ref{eq:line2})]. We ignore the difference of the current
times when we calculate gamma-ray spectra.}. CRs with $p>p_{\rm esc}$
have escaped from the SNR, and their distribution function is given by
equation~(\ref{eq:fext}). The momentum spectrum for the CRs with
$p<p_{\rm esc}$ that have seeped into the cloud is given by $f_{\rm gas}
N_{\rm sh}/2$ if cooling is ignored. The factor of two comes from the
assumption that the CRs that were originally in the overlapped region
(the shaded region in figure~\ref{fig:MeVCRs}a) when $d_{\rm MC}=W_{\rm
sh}/2$ was satisfied have escaped into the cloud
(section~\ref{sec:MeV}). We calculate the gamma-ray spectra using a
model by \citet{2006ApJ...647..692K} and \citet{2008ApJ...674..278K}.

\section{Results}
\label{sec:results}

We apply our model to the SNRs W28 and W44. We chose these
objects because both neutral iron line emissions and gamma-ray emissions
have been detected \citep{2018ApJ...854...87N}. In particular, detailed
gamma-ray spectra are available for these objects. Our procedure is as
follows. (1) Since the distribution and the mass of molecular clouds are
determined mainly through radio observations, we fix the parameters for
the clouds at their observed values (table~\ref{tab:mol}). (2) Using the
observed gamma-ray spectra, we constrain our model parameters for CRs
through $\chi^2$ fitting (table~\ref{tab:para}). (3) From the fitting
results, we estimate the intensities of neutral iron line emissions
(table~\ref{tab:results}) and compare them with the observed
intensities.

%Since our model is rather simple, we just
%show that the observations can be reproduced by using reasonable
%parameters and we do not perform parameter searches.

\subsection{W28}

W28 is a middle-aged SNR from which gamma-rays have been observed in the
GeV \citep{2010ApJ...718..348A,2014ApJ...786..145H,2018ApJ...860...69C}
and the TeV bands \citep{2008A&A...481..401A}. Since previous studies
have shown that the distance to the SNR is $d\sim 2$~kpc
(e.g. \cite{1976ApSS..40...91G,2002AJ....124.2145V}), we assume that
$d=2$~kpc in this study (table~\ref{tab:mol}).

We focus on the northern gamma-ray component (HESS~J1801-233;
\cite{2008A&A...481..401A}), which appears to be associated with the
neutral iron line emissions \citep{2018ApJ...854...87N}. For W28, we
assume that $L_1=12$~pc, $L_2=15$~pc and $f_{\rm gas}=0.1$
(table~\ref{tab:mol}). They are estimated from the distribution
of molecular gas and the gamma-ray images (figure~2 in
\cite{2008A&A...481..401A}), assuming that the cloud is rather
spherical.  We fix the mass of the molecular cloud at the observed value
($M_{\rm gas}\sim 5\times 10^4\: M_\odot$;
\cite{2008A&A...481..401A}). For these parameters, the gas number
density is $n_{\rm H}=3000\:\rm cm^{-3}$.
The current time is $t_{\rm obs}=1.5\times 10^4$~yr
[equation~(\ref{eq:tobs}) and table~\ref{tab:results}].

The gamma-ray spectrum is sensitive to the total CR energy
($E_{\rm tot,CR}$), the maximum momentum of CRs ($p_{\rm max}$), and the
index of the CR energy spectrum ($s$) at the shock front. Thus, we fit
the observed spectrum with our model by varying these three parameters;
the other parameters are fixed. The results are shown in
table~\ref{tab:para}. The escape momentum $p_{\rm esc}$ and the iron
line intensity $I_{\rm 6.4keV}$ can be obtained as a result of the fit
(table~\ref{tab:results}). Figure~\ref{fig:w28} shows that the best-fit
model well reproduces the Fermi
\citep{2010ApJ...718..348A,2018ApJ...860...69C} and the HESS
observations \citep{2008A&A...481..401A}. The iron line intensity is
$I_{\rm 6.4keV}=0.07^{+0.01}_{-0.05}\rm\: photons\: s^{-1}\: cm^{-2}\:
sr^{-1}$ (table~\ref{tab:results}), which is consistent with $I_{\rm
6.4keV}=0.10\pm 0.05\rm\: photons\:\: s^{-1}\: cm^{-2}\: sr^{-1}$
obtained by
\citet{2018ApJ...854...87N}\footnote{\citet{2018ApJ...854...87N} did not
represent $I_{\rm 6.4keV}$ for individual SNRs in their paper.}. On the
other hand, \citet{2018PASJ...70...35O} observed more outside regions
(closer to the rim) of the SNR. Their obtained values of $I_{\rm
6.4keV}$ are generally larger than that reported by
\citet{2018ApJ...854...87N}. In particular, for the region where the
shock is interacting with clouds or the rim of the SNR (region~1 in
their paper), the intensity is $I_{\rm 6.4keV}=0.48\pm 0.28\rm\:
photons\:\: s^{-1}\: cm^{-2}\: sr^{-1}$ if the contribution from the
Galactic ridge X-ray emission is subtracted. This is probably because
the angle between the line of sight and the radial direction of the SNR
is close to $\theta=90^\circ$ (figure~\ref{fig:MeVCRs}b).

\subsection{W44}
\label{sec:w44}

W44 is another middle-aged SNR from which gamma-rays have been observed
in the GeV band; the decrement below $\sim 200$~MeV suggests a hadronic
origin
\citep{2010Sci...327.1103A,2013Sci...339..807A,2014A&A...565A..74C}. On
the other hand, TeV gamma-rays have not been detected
\citep{1998A&A...329..639B,2002A&A...395..803A}. Since the
distance has been estimated to be $d\sim 3$~kpc
\citep{1975A&A....45..239C,1991ApJ...372L..99W}, we assume that
$d=3$~kpc.  The mass of cold gas directly associated with W44 is rather
uncertain. \citet{2013ApJ...768..179Y} estimated that the total
molecular mass around W44 is $\sim 4\times 10^5\: M_\odot$. However, the
radio and gamma-ray images show that only part of the molecular gas
seems to be responsible for the gamma-ray emissions (their figures~3
and~7). From the overlapped area of the radio and gamma-ray emitting
regions, we assume that $M_{\rm gas}= 4\times 10^4\: M_\odot$ and
$f_{\rm gas}=0.1$ (table~\ref{tab:mol}), which result in $n_{\rm
H}=2400\rm\: cm^{-3}$. The mass is larger than that of the shocked gas
($\sim 7\times 10^3\: M_\odot$; \cite{2013ApJ...768..179Y}), which
should be the minimum mass of the gas directly associated with the
SNR. We also assume that $L_1=12$~pc and $L_2=15$~pc by reference to the
gamma-ray image (table~\ref{tab:mol}). The uncertainties of the
parameters are discussed in the next section.
The current time is $t_{\rm obs}=1.5\times 10^4$~yr
(table~\ref{tab:results}).

\begin{figure}
 \begin{center}
  \includegraphics[width=8.0cm]{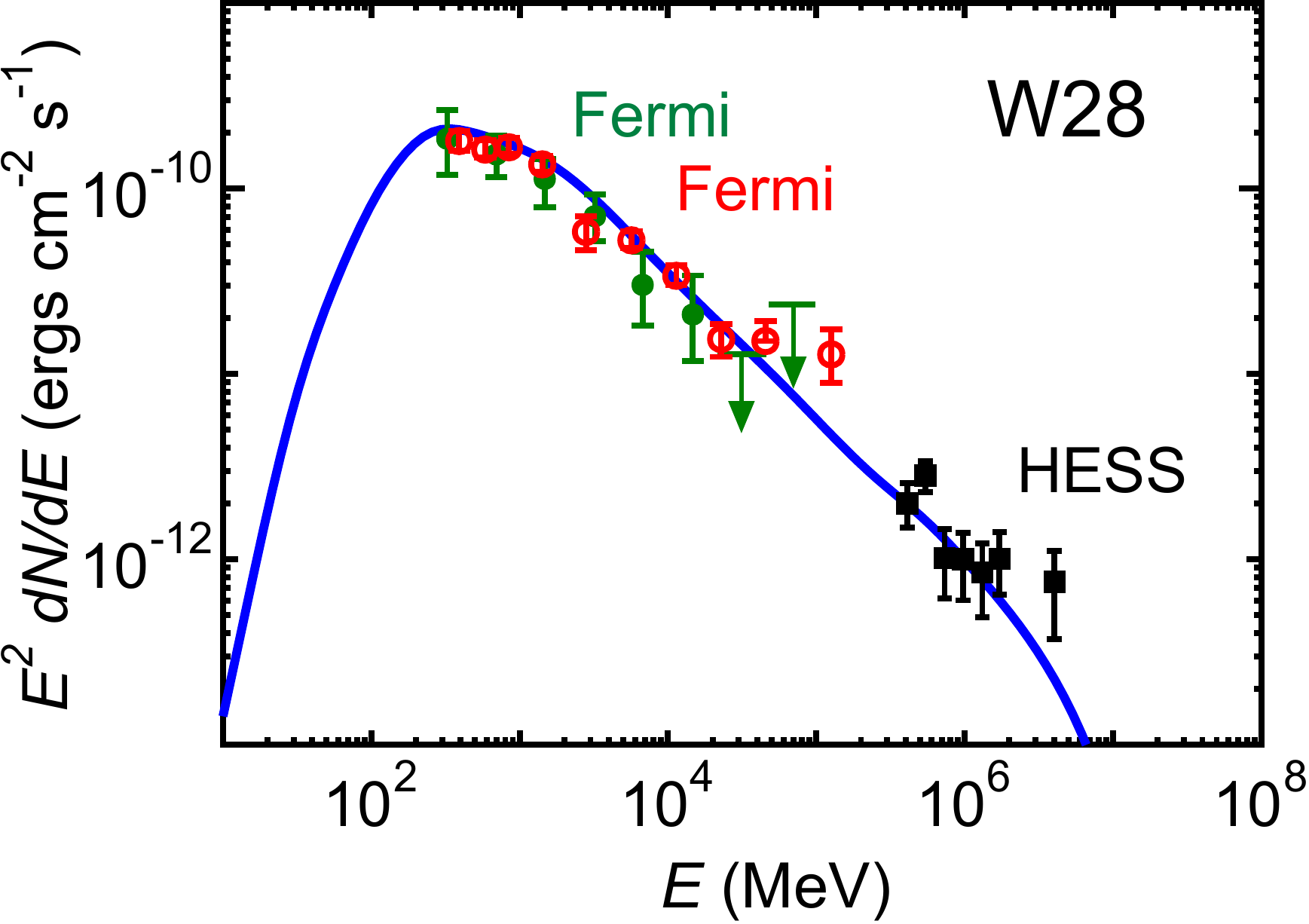}
 \end{center}
 \caption{Comparison of the best-fit model (solid line) with the Fermi
(green filled circles; \cite{2010ApJ...718..348A}, red open circles;
\cite{2018ApJ...860...69C}) and the HESS (black filled squares;
\cite{2008A&A...481..401A}) observations for the SNR
W28.}\label{fig:w28}
\end{figure}

\begin{figure}
 \begin{center}
  \includegraphics[width=8.0cm]{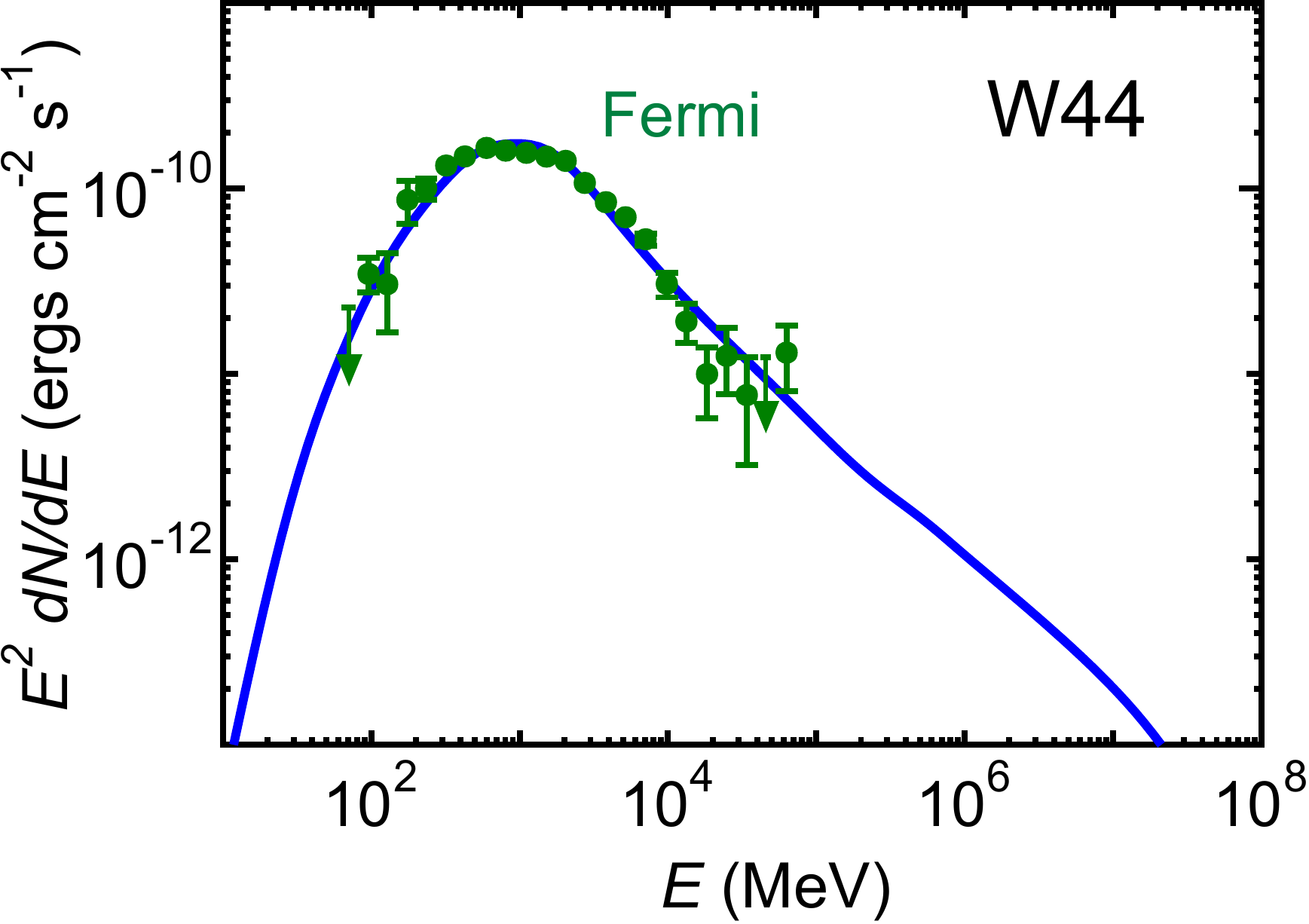}
 \end{center}
 \caption{Comparison of the best-fit model (solid line) with the Fermi
(green filled circles; \cite{2013Sci...339..807A}) observations for the
SNR W44.}\label{fig:w44}
\end{figure}

In figure~\ref{fig:w44}, we present the gamma-ray spectrum of W44. Our
model results are consistent with the Fermi results
\citep{2013Sci...339..807A}.  The peak of the spectrum is attributed to
the break of the CR momentum spectrum at $p=p_{\rm esc}$
[equations~(\ref{eq:Nesc}) and~(\ref{eq:Nsh}); see also
\cite{2011MNRAS.410.1577O}]. The gamma-ray energy at the peak is larger
than that of W28 (figure~\ref{fig:w28}), which reflects the larger value
of $p_{\rm esc}$ (table~\ref{tab:results}). The iron line
intensity is $I_{\rm 6.4keV}=0.10^{+0.04}_{-0.02}\rm\:\: photons\:\:
s^{-1}\: cm^{-2}\: sr^{-1}$ (table~\ref{tab:results}), which is
consistent with $I_{\rm 6.4keV}=0.15\pm 0.08\rm\: photons\:\: s^{-1}\:
cm^{-2}\: sr^{-1}$ obtained by \citet{2018ApJ...854...87N}.

\begin{table}
  \tbl{Input parameters for molecular clouds}{%
  \begin{tabular}{lll}
& & \\
 \hline
 Parameters    & W28 & W44 \\ \hline
   $L_1$ (pc) & 12 & 12 \\
   $L_2$ (pc) & 15 & 15 \\
   $M_{\rm gas}$ ($10^4\: M_\odot$)& 5 & 4 \\
   $f_{\rm gas}$ & 0.1 & 0.1 \\
   $d$ (kpc)& 2 & 3 \\ \hline
  \end{tabular}}\label{tab:mol}
\end{table}

\begin{table}
  \tbl{Fitting results}{%
  \begin{tabular}{lll}
& & \\
 \hline
 Parameters    & W28 & W44 \\ \hline
   $E_{\rm CR,tot}$~($10^{50}$~erg) & $1.8^{+0.2}_{-0.4}$ 
& $3.7^{+0.3}_{-0.4}$ \\
   $p_{\rm max}c$~(TeV) & $40^{+118}_{-26}$ 
& $263^{+25}_{-23}$ \\
   $s$ & $2.02^{+0.03}_{-0.09}$ & $1.98^{+0.12}_{-0.06}$ \\ \hline
  \end{tabular}}\label{tab:para}
\end{table}

\begin{table}
  \tbl{Output parameters}{%
  \begin{tabular}{lll}
& & \\
 \hline
   Parameters & W28 & W44 \\ \hline
   $t_{\rm obs}$ ($10^4$~yr) & 1.5 & 1.5 \\
   $t_{\rm pass}$ ($10^4$~yr) & 0.3 & 0.3 \\
   $u_{\rm sh}$ ($\rm\: km\: s^{-1}$) & 304 & 304 \\
   $p_{\rm esc}c$ (GeV) & $0.6^{+1.9}_{-0.4}$ & $4.1^{+0.4}_{-0.3}$ \\
   $I_{\rm 6.4keV}^*$ (photons~$\rm s^{-1}\: cm^{-2}\: sr^{-1}$)
   & $0.07^{+0.01}_{-0.05}$ & $0.10^{+0.04}_{-0.02}$\\
 \hline
 \end{tabular}}\label{tab:results}
   \begin{tabnote}
 \footnotemark[$*$] The values are for $\theta=0$.
  \end{tabnote}
\end{table}

\section{Discussion}
\label{sec:disc}

We have shown that observational results of W28 and W44 can be explained
by an CR escaping scenario for SNRs. In our model, the SNRs that show
neutral iron line emissions are interacting with surrounding molecular
clouds. While CRs with $E\gtrsim$~GeV have already escaped from the
SNRs, those with $E\sim$~MeV are now leaking into the clouds and
producing the iron line emissions. 

Our model constrains parameters for an SNR by the iron line and
gamma-ray emissions, respectively.  From the iron line intensity $I_{\rm
6.4keV}$, the product of $n_{\rm CR,sh}(E\sim 10~{\rm MeV}) u_{\rm
sh}$ can be obtained from equation~(\ref{eq:line2}). Note that since the
ionization cooling time satisfies $t_{\rm cool}n_{\rm H}=$~const
(e.g. \cite{1994A&A...286..983M}), the iron line intensity is not
dependent of $n_{\rm H}$ as long as $t_{\rm cool}<0.5\: t_{\rm pass}$,
which is fulfilled when $E$ is relatively small ($E\lesssim
100$~MeV). On the other hand, observations of the gamma-ray luminosity
determines the product of $n_{\rm CR}(\gtrsim {\rm GeV}) M_{\rm
gas}$ because the clouds are the targets of CRs in $pp$-interaction.

The parameters regarding molecular clouds, especially for W44
(section~\ref{sec:w44}), have some uncertainties, and there is a kind of
degeneracy in terms of gamma-ray luminosity. For example, if $M_{\rm
gas}$ is doubled by doubling the gas density $n_{\rm H}$, the total CR
energy $E_{\rm CR,tot}$ needs to be halved.  This is because the CR
density $n_{\rm CR}$ needs to be halved to be consistent with the
observed gamma-ray luminosity. In this case, $I_{\rm 6.4keV}$ is reduced
but it is not exactly halved because of the contribution of CRs with
$E\gtrsim 100$~MeV for which the relation of $t_{\rm cool}n_{\rm
H}=$~const is not valid. The intensity $I_{\rm 6.4keV}$ is marginally
consistent with observations for W28 and W44. Similarly, if $M_{\rm
gas}$ is doubled by doubling the filling factor $f_{\rm gas}$, $E_{\rm
CR,tot}$ needs to be halved. As a result, the CR density $n_{\rm CR}$ is
halved, and $I_{\rm 6.4keV}$ is halved. These could be used to verify
our model by future observations of cold gas. For example, given the
currently observed gamma-ray luminosity and $I_{\rm 6.4keV}$, it is
unlikely that $M_{\rm gas}$ is larger than those we assumed by an order
of magnitude, although there is an uncertainty about the dependence of
$I_{\rm 6.4keV}$ on the angle $\cos\theta$ (section~\ref{sec:MeV}).  The
lower limit of $M_{\rm gas}$ can be constrained by the upper limit of
$E_{\rm CR,tot}$ because $E_{\rm CR,tot}$ cannot exceed the explosion
energy of an supernova ($\sim 10^{51}$~erg). Thus, it is unlikely that
$M_{\rm gas}$ is smaller than those we assumed by an order of
magnitude.

For $n_{\rm H}>1000\rm\: cm^{-3}$, the cooling time of CRs is very short
($t_{\rm cool}\lesssim 100$~yr at $E\sim 10$~MeV). This means that MeV
CRs that enter the cloud almost immediately lose their energy. Thus, the
neutral iron line emissions can be observed in a time scale of $\sim
t_{\rm pass}$. Table~\ref{tab:results} shows that $t_{\rm pass}$ is 20\%
of the current time $t_{\rm obs}$ and thus the duration is not extremely
small compared with the age of the SNRs. In other words, the possibility
of observing the iron line emissions is not tiny. Moreover, if multiple
clouds are randomly located around the SNR, the iron line emissions
could blink on and off as the shock front passes the clouds. We note
that we assumed that the fraction of the confinement region ($W_{\rm
sh}/R_{\rm sh}\propto \kappa$) for MeV CRs is the same as that for
GeV--TeV CRs. If this is not the case, the duration $t_{\rm pass}$ and
the line intensity $I_{\rm 6.4keV}$ can change
[equations~(\ref{eq:nCRsh}) and~(\ref{eq:line2})]. For example, if
$W_{\rm sh}$ is doubled, $t_{\rm pass}$ is doubled and $I_{\rm 6.4keV}$
is halved.

\section{Conclusion}
\label{sec:conc}

We have shown that 6.4keV neutral iron line emissions and gamma-ray
emissions from SNRs can be explained by an CR escaping scenario for
SNRs. In this model, the SNRs with the iron line emissions are
interacting with surrounding molecular clouds. We assume that CRs are
accelerated at the SNR with a single power-law spectrum. When the SNR
comes into contact with the clouds, MeV CRs are still confined in the
SNR. They gradually leak into the clouds and produce the iron line
emissions through interaction with irons in the clouds. On the the hand,
the CRs with $E\gtrsim$~GeV have already escaped from the SNR at the
contact.

We applied this model to the SNRs W28 and W44 and showed that both the
observed iron line intensities and the gamma-ray spectra can be
reproduced. These support a hadronic scenario for the gamma-ray
emissions from the SNRs.

\begin{ack}
We thank the anonymous referee, whose comments improved the clarity of
this paper. This work was supported by MEXT KAKENHI No.~18K03647
(Y.F.), JP16J00548 (K.K.N.), JP15H02070 (H.M.), and 16K17702 (Y.O.).
\end{ack}

%\bibliography{paper}
%\bibliographystyle{junsrt}

\end{document}